\documentclass[aps,prmaterials,reprint,superscriptaddress,nofootinbib,amsmath,amssymb]{revtex4-1}

\usepackage{times}
\usepackage{graphicx}
\usepackage{booktabs}
\usepackage{gensymb}
\usepackage{xcolor}
\usepackage[colorlinks]{hyperref}

\newcommand{\germania}{GeO$_2$}
\newcommand{\silica}{SiO$_2$}
\newcommand{\tantala}{Ta$_2$O$_5$}
\newcommand{\titania}{TiO$_2$}
\newcommand{\tdg}{TiO$_2$-doped GeO$_2$}

\begin{document}

\preprint{APS/123-QED}

\title{Atomic Structure of Amorphous Optical Coatings of \titania-doped \germania~by Grazing-Incidence Total X-ray Scattering Measurements}

\author{K. Prasai}
\email{kprasai@kennesaw.edu}
\affiliation{Department of Physics, Kennesaw State University, Marietta, Georgia 30060, USA}

\author{J. Jiang}
\thanks{K. Prasai and J. Jiang contributed equally to this work.}
\affiliation{Department of Physics, Northeastern University, Boston, Massachusetts 02115, USA}

\author{K.\,Lee}
\affiliation{Department of Physics, Sungkyunkwan University, Seoul 03063, Republic of Korea}

\author{L.\,Yang}
\affiliation{Department of Electrical and Computer Engineering, Colorado State University, Fort Collins, Colorado 80523, USA}

\author{M.\,Chicoine}
\affiliation{Department of Physics, Universit\'e de Montr\'eal,  Montréal, Qu\'ebec H3T 1J4, Canada}

\author{S.\,Khadka}
\affiliation{Department of Physics, Kennesaw State University, Marietta, GA 30060, USA}

\author{A.\,Markosyan}
\affiliation{E. L. Ginzton Laboratory, Stanford University, Stanford, California 94305, USA}

\author{A.\,Mehta}
\affiliation{SLAC National Accelerator Laboratory, Menlo Park, California 94025, USA}

\author{C.\,S.\,Menoni}
\affiliation{Department of Electrical and Computer Engineering, Colorado State University, Fort Collins, Colorado 80523, USA}

\author{S.\,Patel}
\affiliation{Department of Physics, Kennesaw State University, Marietta, GA 30060, USA}

\author{F.\,Schiettekatte}
\affiliation{Department of Physics, Universit\'e de Montr\'eal,  Montréal, Qu\'ebec H3T 1J4, Canada}

\author{B.\,Shyam}
\affiliation{Xerion Advanced Battery Corp., Kettering, Ohio 45420, USA}

\author{G.\,Vajente}
\affiliation{LIGO Laboratory, California Institute of Technology, Pasadena, California 91125, USA}

\author{H-P.\,Cheng}
\affiliation{Department of Physics, Northeastern University, Boston, Massachusetts 02115, USA}

\author{M.\,M.\,Fejer}
\affiliation{E. L. Ginzton Laboratory, Stanford University, Stanford, California 94305, USA}

\author{R.\,Bassiri}
\affiliation{E. L. Ginzton Laboratory, Stanford University, Stanford, California 94305, USA}

\date{\today}

\begin{abstract}
\noindent Reducing coating thermal noise in future gravitational-wave detectors requires identifying the atomic motifs that control mechanical loss in amorphous optical coatings. We combine grazing-incidence X-ray pair distribution function measurements with atomic-structure modeling to study amorphous TiO$_2$-doped GeO$_2$ films over Ti cation concentrations from $\sim$11\% to $\sim$48\%, before and after annealing. The structural analysis reveals systematic composition- and annealing-dependent changes in short- and intermediate-range order. Increasing Ti content raises the average Ti coordination and promotes edge- and face-sharing polyhedral connections, while Ge remains predominantly fourfold coordinated. Annealing reduces these compact shared-polyhedron motifs and sharpens the first sharp diffraction peak, indicating a more relaxed intermediate-range network. Among the structural descriptors examined, the clearest correlation with the annealing-induced reduction in mechanical loss is the decrease in edge- and face-sharing polyhedra. These results connect composition, annealing, atomic structure, and mechanical dissipation in TiO$_2$-doped GeO$_2$, providing microscopic guidance for optimizing low-noise mirror coatings.
\end{abstract}

\pacs{Valid PACS appear here}

\maketitle

\section{INTRODUCTION}
The current generation of interferometric gravitational wave detectors, such as Advanced LIGO and Virgo, are limited in sensitivity by various noise sources, including coating thermal noise (CTN) originating from the mirror coatings used to reflect the laser beams \cite{martynov2016sensitivity, buikema2020sensitivity}. These coatings are dielectric mirrors consisting of alternating layers of ion-beam-sputtered amorphous \titania-doped \tantala~and \silica~\cite{granata2020amorphous, amato2019optical}. Planned future detectors, such as Advanced LIGO+, Cosmic Explorer or Einstein Telescope, will require a reduction of CTN in order to meet their target sensitivities \cite{punturo2010einstein}.

It has previously been established that in oxide glasses the mechanical loss is associated with two-level systems (TLS) present in the atomic structure of the coating materials. In particular, specific motifs in the atomic structure appear to correlate with mechanical loss; for example, room-temperature mechanical losses are connected to face- and edge-sharing polyhedral motifs, while cryogenic losses are connected with corner-sharing polyhedra \cite{prasai2019high}. On the basis of this insight, it was predicted and subsequently empirically verified that GeO$_2$-based glasses are a suitable base material for low-noise, high-index layers, and that Ti doping yields Ti:GeO$_2$ with adequately high refractive index and low mechanical loss to reduce CTN by approximately a factor of two \cite{vajente2021low}.

Mechanical loss in \tdg~films has been observed to depend on \titania~concentration \cite{vajente2021low}, with undoped \germania~films exhibiting lower mechanical loss than doped films. However, compared to undoped \germania~films, Ti‑doped layers offer a higher refractive‑index contrast with the low‑index silica layers, thereby reducing the number of coating layers required to achieve the target reflectivity. Because CTN decreases with decreasing layer thickness, there exists an optimal Ti doping level that balances reduced coating thickness with minimized mechanical loss, yielding the lowest overall CTN. Therefore, one route toward further reducing CTN is to identify an optimized \tdg~coating composition with significantly lower mechanical loss and/or increased refractive index. For amorphous oxide films prepared by physical vapor deposition, additional processing parameters—such as substrate temperature, deposition rate, and ion energies during deposition—can also significantly influence mechanical loss \cite{vajente2018effect, granata2020amorphous, yang2021enhanced}. Post‑deposition annealing, performed below the crystallization temperature of the films, has been shown to consistently reduce mechanical loss \cite{prasai2019high,vajente2021low, abernathy2021exploration}. A clearer understanding of how these factors influence mechanical loss can be achieved by studying the atomic‑scale structure of \tdg~films.

In this paper, we present a detailed study of the atomic structure of \tdg~films at four \titania~doping levels. We report results from X-ray grazing-incidence pair distribution function (GIPDF) measurements and analyze atomic models refined against these measurements. A goal of this paper is to complement earlier mechanical-loss measurements on \tdg~films \cite{vajente2021low, khadka2023cryogenic} and to aid in designing coatings with further reduced CTN.

\section{METHODS} \label{sec:methods}

\subsection{Film Deposition and Characterization} 
Amorphous films of \tdg~were deposited on 1-mm-thick, 75-mm-diameter silica disks using the ion-beam sputtering (IBS) method described in \cite{vajente2021low,yang2021enhanced}. Multiple samples were deposited at each nominal Ti-cation concentration of approximately 11\%, 27\%, 40\%, and 48\%. Two samples at each of these compositions were taken for GIPDF measurements. One of the two samples at each composition was annealed in air at 500$\degree$C for 10 hours. Rutherford backscattering spectrometry (RBS) was used to measure the cation concentration, oxygen stoichiometry, and atomic areal density of the films \cite{chu2012backscattering}. Thickness was measured using spectroscopic ellipsometry \cite{fujiwara2007spectroscopic}. Mass density was computed from the RBS and ellipsometry results. Mechanical-loss measurements were carried out using the gentle nodal suspension (GeNS) method \cite{cesarini2009gentle, vajente2017high}. These results are shown in Table~\ref{table:1}. Details of the IBS, RBS, ellipsometry, and GeNS instrumentation, and their application to similar samples, can be found in Ref. \cite{lalande2024ar} for similar samples.

\subsection{GIPDF Measurements} 
GIPDF data were collected at the dedicated X-ray scattering beamline 10-2 at the Stanford Synchrotron Radiation Lightsource (SSRL). Using an energy-resolved point detector to scan over scattering angles, a $q$-dependent scattering profile was collected for each sample. A grazing-incidence angle was chosen to preferentially collect X-rays scattered from the coatings (rather than the substrate) while maintaining a $q$-range up to 20.1\,\AA$^{-1}$. The X-ray energy was 21.5~keV. The elastic signal from scattered X-rays was used to obtain the total scattering signal from the coating, while the fluorescence signal from the sample was used to correct for the detector footprint in the measured intensities. The total scattering data were reduced to the normalized structure factor $S(q)$ after applying identical corrections to all samples for air scattering, absorption, Compton scattering, and polarization effects. The structure factors were then Fourier-transformed to $r$-space to obtain the X-ray PDFs for each sample. Further details of the GIPDF data-collection method are discussed in Refs.~\cite{shyam2016measurement,qiu2004pdfgetx2}. For the 48\% Ti samples, GIPDF data were collected only for the annealed film due to time constraints.

\begin{table*}[t]
  \caption{Film properties: Measured properties of the films including atomic composition from RBS measurement, thickness and mass density from spectroscopic ellipsometry and RBS measurements, and mechanical loss from GeNS measurements. Expected uncertainties in mass densities are $\pm 0.02$ to $\pm 0.03$ g/cm$^3$. RBS found that, within the uncertainty of the method ($\sim$2\%), the samples were stoichiometric in O content. Reported mechanical-loss values ($\phi$) for annealed samples are from nominally identical samples from the same deposition run. The latter measurements were carried out at room temperature at kHz frequencies (modes between 1 kHz to 31 kHz).}
  \label{table:1}
  \centering
  \setlength{\tabcolsep}{5pt} 
  \renewcommand{\arraystretch}{1.05} 
  \begin{tabular*}{\textwidth}{@{\extracolsep{\fill}}ccccccccc}
      \specialrule{.08em}{0.3em}{0.3em}
      Ti$_{\text{nominal}}$ & $T_{\text{anneal}}$ & \multicolumn{3}{c}{RBS (\%)} & Ti$_{\text{actual}}$ & $\rho$ & $t$ & $\phi$ \\
      (cat.\ \%) & ($^\circ$C) & Ar & Ti & Ge & (cat.\ \%) & (g/cm$^3$) & (nm) & ($\times 10^{-4}$) \\
      \specialrule{.05em}{0.2em}{0.2em}
      11 & AD  & 1.9 & 3.4  & 28.0 & 10.8$\pm$0.5 & 3.49 & 574 & 3.47$\pm$0.25 \\
         & 500 & 0.0 & 3.9  & 29.4 & 11.7$\pm$0.5 & 3.43 & 613 & 2.37$\pm$0.24 \\
      \specialrule{0em}{0.1em}{0.23em}
      27 & AD  & 1.7 & 8.5  & 23.1 & 26.9$\pm$0.5 & 3.56 & 480 & 4.81$\pm$0.45 \\
         & 500 & 0.0 & 9.4  & 23.8 & 28.3$\pm$0.5 & 3.56 & 515 & 1.53$\pm$0.23 \\
      \specialrule{0em}{0.1em}{0.23em}
      40 & AD  & 1.7 & 12.6 & 19.0 & 39.9$\pm$0.5 & 3.73 & 598 & 4.02$\pm$0.24 \\
         & 500 & 0.5 & 13.4 & 19.4 & 40.8$\pm$0.5 & 3.62 & 607 & 1.80$\pm$0.08 \\
      \specialrule{0em}{0.1em}{0.23em}
      48 & AD  & 1.7 & 15.1 & 16.5 & 47.8$\pm$0.5 & 3.55 & 518 & 5.69$\pm$0.85 \\
         & 500 & 1.0 & 15.7 & 16.6 & 48.6$\pm$0.5 & 3.63 & 523 & 2.64$\pm$0.32 \\
      \specialrule{.08em}{0em}{0em}
  \end{tabular*}
\end{table*}

\subsection{Generating Structure Models}
We generated atomic models of amorphous \tdg~with a focus on capturing the structure of the films for which mechanical-loss and GIPDF data were collected. In such cases, regression methods such as inverse simulated annealing \cite{los2013inverse} or reverse Monte Carlo (RMC) \cite{keen1990structural} are commonly used to fit atomic coordinates by minimizing differences between measured PDFs and those computed from models. However, unaided regression algorithms are known to produce nonphysical structural solutions even in elemental systems \cite{drabold2009topics}. Here we use the Force Enhanced Atomic Refinement (FEAR) method \cite{pandey2015force,pandey2016inversion}, which iterates a few hundred steps of RMC with a few steps of energy minimization \cite{guenole2020assessment} until convergence in $\chi^2$ and energy is obtained. We define
\begin{equation}\label{eq1}
	\chi^2 = \sum_j \frac{\left(S_{j}^{\text{expt}}(q)-S_{j}^{\text{calc}}(q)\right)^2}{\sigma_{j}^{2}},
\end{equation}
\noindent where the sum is over $j$ indexing discrete bins in the $q$-range of interest, $S^{\text{expt}}(q)$ and $S^{\text{calc}}(q)$ are the experimental and modeled structure factors, respectively, and $\sigma_{j}$ is a weighting factor commonly used to represent $q$-dependent noise. In this work, we set $\sigma_j=1.0$ for all $q$.

\begin{figure}[h]
	\centering
	\includegraphics[width=\linewidth]{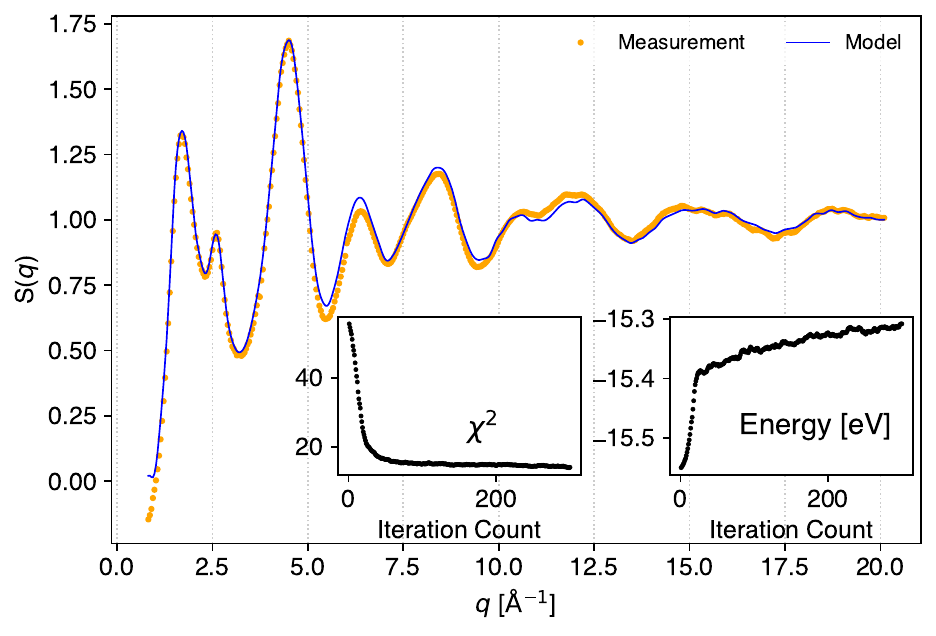}
	\caption{Structure factor $S(q)$ from measurement and modeling for the 11\% Ti unannealed sample. Insets show $\chi^{2}$ and average energy at each iteration of FEAR modeling. The $S(q)$ curve in the main panel corresponds to the final (300$^{\textrm{th}}$) iteration.}
	\label{fig:goodness-of-fit}
\end{figure}

Atomic models were generated for each available GIPDF dataset, giving seven modeled datasets in total because the 48\% Ti as-deposited GIPDF dataset was not available. For each dataset, we first generated random atomic configurations in a periodic simulation cell whose composition and mass density matched the corresponding measurements in Table~\ref{table:1}. Since RBS detected Ar in several coatings, with concentrations up to 1.9\%, the corresponding concentration of Ar atoms was included explicitly in the models. Although Ar is not part of the oxide network, it can occupy voids in ion-beam-sputtered films and has a finite X-ray scattering contribution. The random configurations were converted into amorphous starting structures by quench-from-the-melt molecular dynamics using a BKS--Morse potential with an added three-body term \cite{jiang2025machine}. Ar--Ar interactions were modeled using a Lennard--Jones 6--12 potential \cite{bernardes1958theory} whereas Ar--host interactions were modeled using the ZBL potential \cite{ziegler1985stopping}. The force-field form and parameters are given in Appendix \ref{app:potential}.

The quenched structures were then used as starting configurations for FEAR modeling. In each FEAR cycle, reverse Monte Carlo moves were used to reduce the difference between the measured and calculated structure factors through the \(\chi^2\) defined in Eq.~(\ref{eq1}), and short energy-minimization steps were carried out using the same force field described above. A bond-valence-sum penalty was included in the effective RMC cost function, and RMC moves were constrained so that all metal--oxygen bond distances remained at least 1.5~\AA. For each dataset, the FEAR procedure was repeated for 100 independent starting configurations and random seeds, producing 100 statistically independent models per sample.

The final models, by construction, agree well with the measured X-ray data; an example is shown in Fig.~\ref{fig:goodness-of-fit}. The insets show that $\chi^2$ decreases rapidly during the early stages of FEAR refinement and then approaches a plateau. Over the same iterations, the empirical potential energy increases. This increase is expected because the melt--quench configurations used as FEAR starting points are low-energy structures with respect to the approximate classical force field, whereas FEAR minimizes an effective cost function that prioritizes agreement with the measured scattering data while using the force field as a physical regularization. Thus, refinement can move the models away from the empirical force-field minimum when required to reproduce experimentally constrained features of the amorphous network, including cation--cation correlations, polyhedral connectivity, and intermediate-range order. The energy increase should therefore not be interpreted as a loss of physicality, but as the cost of imposing structural information contained in the GIPDF data that is not fully captured by the empirical potential alone.

Both $\chi^2$ and the average energy change most strongly up to $\sim$100 iterations, beyond which only small improvements in $\chi^2$ are obtained along with a slight further increase in energy. For all models presented here, refinement was stopped after the 300$^{\textrm{th}}$ iteration. We verified that the reported structural metrics are unchanged when refinement is stopped at the 400$^{\textrm{th}}$ or 500$^{\textrm{th}}$ iteration.

For each dataset, the modeling program was run 100 times in parallel using independent starting configurations and random seeds, resulting in 100 statistically independent atomic models per sample. In the remainder of this paper, each ensemble of 100 models is referred to by the name of the corresponding sample. All reported quantities are averages over these 100 models, and the error bars represent the corresponding standard deviations.

\section{RESULTS AND DISCUSSION}
\subsection{GIPDF measurements}

\begin{figure*}[t]
	\centering
	\includegraphics[width=\linewidth]{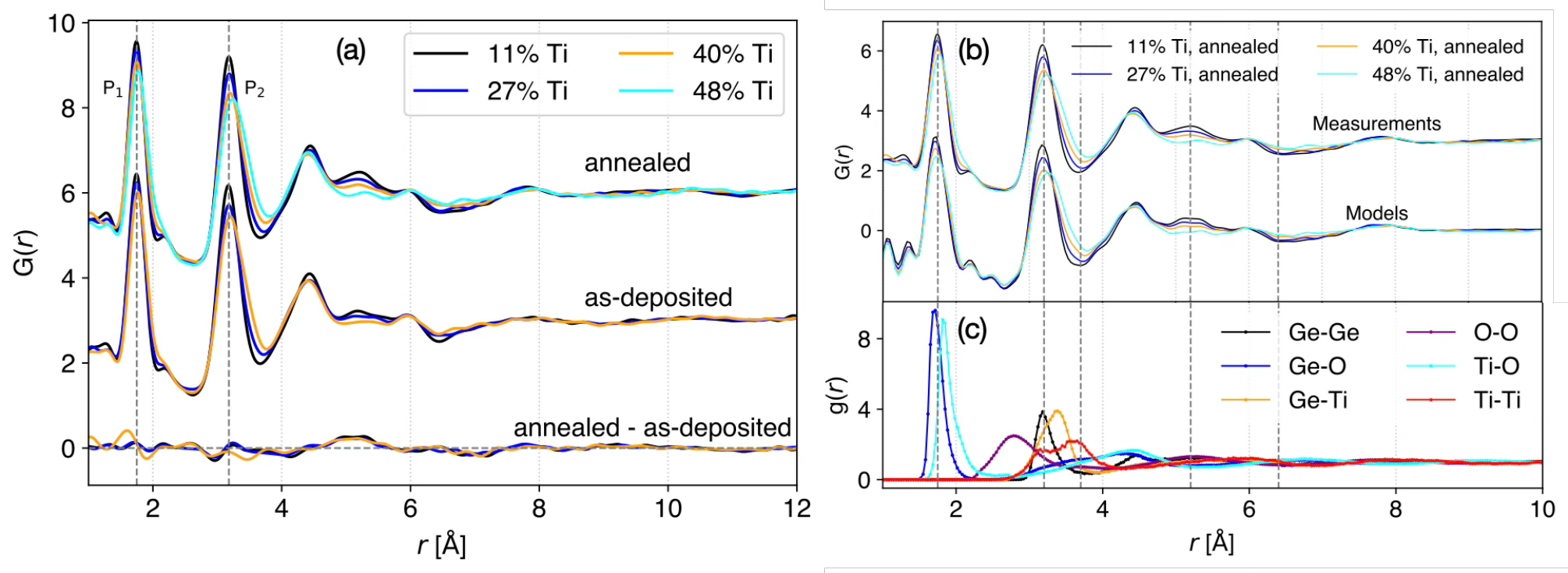}
	\caption{Pair distribution functions obtained from grazing-incidence X-ray total scattering. 
	(a) Measured GIPDFs $G(r)$ from thin films of \tdg. The PDFs of the as-deposited and annealed films are shifted vertically by 3 and 6 units, respectively. The difference between the annealed and as-deposited PDFs is shown near zero on the vertical axis. The first and second main peaks, P$_1$ and P$_2$, occur at 1.75~\AA\ and 3.18~\AA, respectively. The same data-processing protocol was used for all curves, and the observed differences are robust to small changes in the processing parameters. 
	(b) GIPDFs measured from annealed samples together with the corresponding modeled PDFs. With increasing Ti concentration, the GIPDFs show systematic changes, most clearly near $r\sim$1.75~\text{\AA}, 3.2~\text{\AA}, 3.7~\text{\AA}, 5.2~\text{\AA}, and 6.4~\text{\AA}, as indicated by the dashed vertical lines. The modeled PDFs reproduce the measured data well and capture the observed Ti-concentration-dependent structural trends. 
	(c) Number-density-based partial PDFs from the 48\% Ti, annealed models, defined following Eq.~8 of Ref.~\cite{keen2001comparison}.}
	\label{fig:GIPDF}
\end{figure*}

The measured GIPDFs are shown in Fig.~\ref{fig:GIPDF}(a). All samples exhibit a strong first peak, labeled P$_1$, at 1.75~\AA. This peak is assigned primarily to nearest-neighbor Ge--O and Ti--O correlations. With increasing Ti concentration, the right shoulder of P$_1$ becomes broader, consistent with the slightly longer average Ti--O distance compared with the Ge--O distance obtained from the models: 1.82$\pm$0.025~\AA\ for Ti--O and 1.75$\pm$0.025~\AA\ for Ge--O; see Fig.~\ref{fig:GIPDF} (c). The second peak, labeled P$_2$, occurs near 3.18~\AA\ and contains contributions from Ge--Ge, Ge--Ti, and Ti--Ti cation--cation correlations. The increasing Ti content modifies the relative weights of these partial correlations and leads to a broader P$_2$ feature.

Figure~\ref{fig:GIPDF}(a) also shows the difference between the GIPDFs of the as-deposited and annealed films. Near P$_1$ and P$_2$, the annealing-induced differences are more pronounced for the 40\% Ti films than for the 11\% Ti and 27\% Ti films. These differences mainly reflect small annealing-induced shifts and changes in peak shape. The changes near P$_1$ are consistent with changes in Ti coordination, while the Ge coordination remains comparatively insensitive to annealing, as shown in Table~\ref{table:2}. The changes near P$_2$ are consistent with a reduction in edge-sharing (ES) and face-sharing (FS) polyhedral connections and a corresponding increase in corner-sharing (CS) connections after annealing. The larger changes observed for the 40\% Ti films indicate that this composition contains a larger population of fivefold-coordinated Ti atoms and compact ES/FS polyhedral motifs than the 11\% and 27\% Ti films, a trend confirmed by the atomic-structure models discussed below. A further annealing-induced change is visible near 5~\AA, suggesting a change in intermediate-range order (IRO), discussed further in Sec.~\ref{sec:fsdp}.

The GIPDFs also evolve systematically with Ti concentration. For this comparison, we focus on the annealed samples because annealing reduces deposition-related structural variability and provides a more consistent basis for comparing composition-dependent trends. As shown in Fig.~\ref{fig:GIPDF}(b), increasing Ti concentration produces clear changes in the GIPDFs, most prominently near $r\sim$1.75~\text{\AA}, 3.2~\text{\AA}, 3.7~\text{\AA}, 5.2~\text{\AA}, and 6.4~\text{\AA}. These changes are reproduced by the corresponding atomic models, indicating that the models capture the main structural evolution of the films with increasing Ti concentration.

To identify the structural origin of these composition-dependent changes, Fig.~\ref{fig:GIPDF}(c) shows the number-density-based partial PDFs from the 48\% \tdg models. These partial PDFs are used here primarily to identify the characteristic positions of the main pair correlations. The first-neighbor Ge--O and Ti--O peaks occur at 1.75~\AA\ and 1.82~\AA, respectively. Their overlap produces the main P$_1$ feature in the total GIPDF, while the longer Ti--O distance accounts for the increasing broadening of the right shoulder of P$_1$ with increasing Ti concentration. The O--O first-neighbor peak occurs near 2.80~\AA; however, this correlation has only a weak signature in the total X-ray GIPDF because oxygen has a comparatively small X-ray scattering weight.

The strongest Ti-dependent changes occur in the P$_2$ region, which is dominated by cation--cation correlations. In the partial PDFs, the Ge--Ge peak is centered near 3.18~\AA, the Ge--Ti peak near 3.40~\AA, and the Ti--Ti correlation is split into two contributions. The shorter Ti--Ti peak, near 3.15~\AA, is associated with compact ES and FS polyhedral connections, whereas the longer Ti--Ti peak, near 3.60~\AA, is associated with CS polyhedral connections. Consequently, increasing Ti concentration enhances and broadens the cation--cation contribution to P$_2$, especially on its left shoulder, reflecting the growing population of Ti-rich compact polyhedral motifs.

A bond cutoff of 2.6~\AA\ is inferred from the position of first minima of Ge-O and Ti-O partial PDFs and is used throughout the analysis that follows. We have checked that the structure trends reported in this work are not influenced by a slight change ($\pm$0.1~\AA) in bond cutoff distance. 

\begin{figure}[t]
	\centering
	\includegraphics[width=\linewidth]{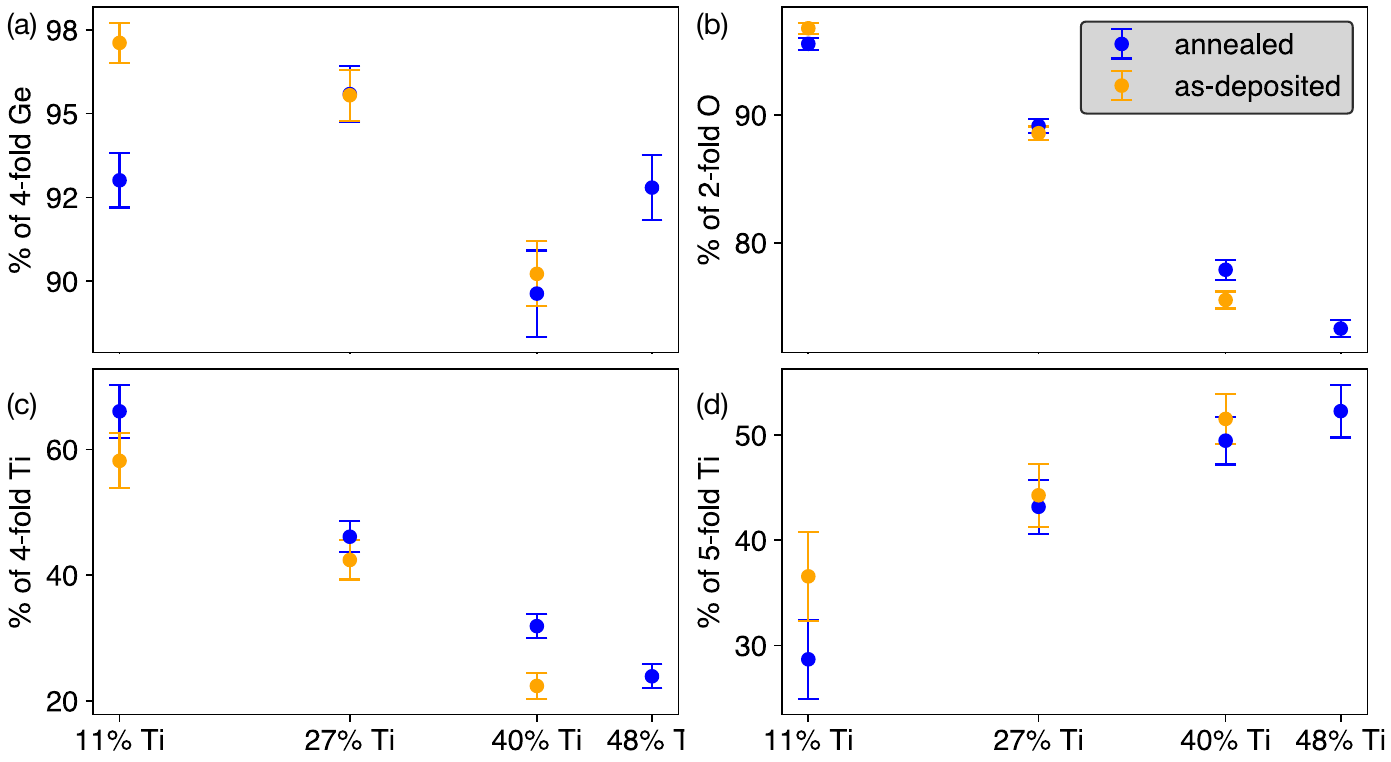}
	\caption{Coordination statistics computed from the experimentally constrained structural models. The panels show the percentages of (a) 4-fold-coordinated Ge atoms, (b) 2-fold-coordinated O atoms, (c) 4-fold-coordinated Ti atoms, and (d) 5-fold-coordinated Ti atoms. Ge remains predominantly 4-fold coordinated across the measured composition range, indicating that the Ge-centered tetrahedral units are largely preserved upon Ti incorporation. In contrast, increasing Ti concentration reduces the fraction of 2-fold-coordinated O atoms and 4-fold-coordinated Ti atoms while increasing the fraction of 5-fold-coordinated Ti atoms. These trends show that Ti incorporation primarily modifies the network through higher-coordination Ti-centered environments and more highly connected oxygen sites.}
	\label{fig:coordination}
\end{figure}

In Fig.~\ref{fig:coordination}, we show the percentages of 4-fold coordinated Ge atoms, 2-fold coordinated O atoms, and 4-fold and 5-fold coordinated Ti atoms. These quantities describe how the local network connectivity changes as Ti is incorporated into the amorphous \germania\ network. In pure amorphous \germania, Ge is expected to occur predominantly in tetrahedral GeO$_4$ units connected through bridging oxygen atoms. The models show that this basic Ge-centered tetrahedral motif is largely preserved in \tdg: the majority of Ge atoms remain 4-fold coordinated across the composition range studied. However, the percentage of 4-fold-coordinated Ge atoms shows a weak non-monotonic decrease with increasing Ti concentration, indicating that Ti incorporation introduces increasing distortion and local chemical disorder into the Ge-centered network.

The oxygen and Ti coordination statistics show a more systematic evolution with Ti content. With increasing Ti concentration, the percentage of 2-fold-coordinated O atoms decreases, indicating that a growing fraction of oxygen atoms participates in more highly connected local environments. This trend is consistent with the fact that Ti in amorphous oxides can adopt higher and more variable coordination than Ge. Correspondingly, the percentage of 4-fold-coordinated Ti atoms decreases while the percentage of 5-fold-coordinated Ti atoms increases. Thus, increasing Ti content does not simply substitute Ti into a fixed tetrahedral \germania-like network; instead, Ti progressively changes the local topology by favoring higher-coordination Ti-centered polyhedra and more highly connected oxygen environments.

Average coordination numbers for Ge, O, and Ti are given in Table~\ref{table:2}. The average Ge coordination remains close to four for all compositions and annealing conditions, whereas the average O and Ti coordination numbers increase with Ti concentration. This contrast indicates that Ge acts as the more structurally conservative network-forming cation, retaining predominantly tetrahedral GeO$_4$ coordination, while Ti accommodates increasing Ti content through a broader distribution of local coordination environments. The increase in $n_{\mathrm{Ti}}$ from the low-Ti to high-Ti films is especially important because higher Ti coordination naturally promotes a denser and more connected network of Ti-centered polyhedra. This change in short-range coordination provides the local structural basis for the composition-dependent changes in the GIPDFs and for the increase in compact polyhedral connections discussed below.

\begin{figure}[t]
	\centering
	\includegraphics[width=\linewidth]{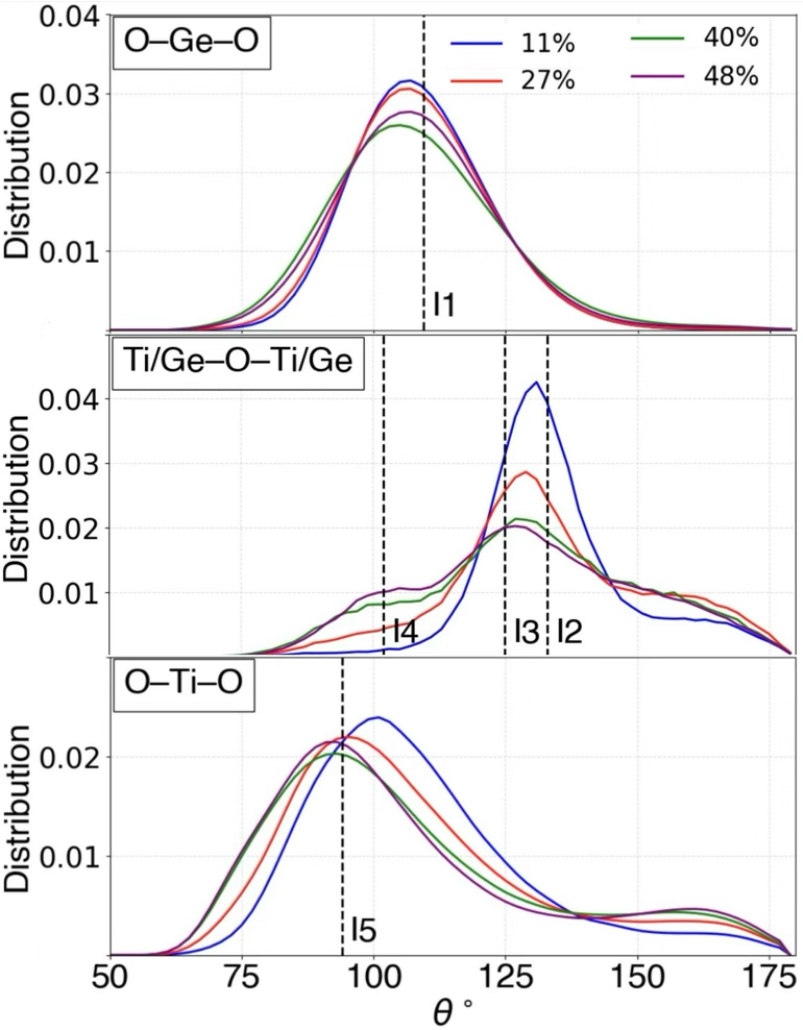}
	\caption{Bond-angle distributions computed from the experimentally constrained structural models. The top panel shows O--Ge--O angles subtended at Ge centers, the middle panel shows cation--O--cation angles subtended at O centers, and the bottom panel shows O--Ti--O angles subtended at Ti centers. Dashed grey lines mark representative peak positions from prior studies of pure amorphous \germania\ and amorphous \titania. From Ref.~\cite{neuefeind1996bond}, $l_1$ is the O--Ge--O peak position at 109.47$\degree$ and $l_2$ is the Ge--O--Ge peak position at 133$\degree$ for amorphous \germania. From Ref.~\cite{triana2016electronic}, $l_3$ and $l_4$ are Ti--O--Ti peak positions at $\sim$102$\degree$ and $\sim$125$\degree$, respectively, for amorphous \titania, and $l_5$ is the O--Ti--O peak position at $\sim$94$\degree$. 
    }
	\label{fig:bondangles}
\end{figure}

Annealing produces smaller changes in the average coordination numbers than changing Ti concentration, but the trends are still physically meaningful. At fixed composition, annealing generally reduces the average Ti coordination slightly, while leaving the Ge coordination nearly unchanged. This suggests that annealing mainly relaxes the more flexible Ti-centered environments rather than substantially reorganizing the Ge-centered tetrahedral backbone. The persistence of nearly fourfold Ge coordination is also consistent with the O--Ge--O bond-angle distributions in Fig.~\ref{fig:bondangles}, which remain centered near the tetrahedral angle across the Ti concentration range, even though the distributions broaden as Ti is added.

\begin{table}[t]
  \caption{Average coordination numbers of Ge, O, and Ti computed from the experimentally constrained models. Error values are standard deviations over the model ensemble. The near-constant value of $n_{\mathrm{Ge}}$ shows that Ge remains predominantly tetrahedral, whereas the increase in $n_{\mathrm{O}}$ and $n_{\mathrm{Ti}}$ with Ti concentration reflects the formation of more highly connected Ti--O environments.}
  \label{table:2}
  \centering
  \setlength{\tabcolsep}{6pt}
  \renewcommand{\arraystretch}{1.05}
  \begin{tabular*}{\columnwidth}{@{\extracolsep{\fill}}ccccc}
      \specialrule{.08em}{0.3em}{0.3em}
      Ti$_{\text{nominal}}$ & $T_{\text{anneal}}$ & $n_{\mathrm{Ge}}$ & $n_{\mathrm{O}}$ & $n_{\mathrm{Ti}}$ \\
      \specialrule{.05em}{0.2em}{0.2em}
      11 & AD  & 3.99$\pm$0.03 & 2.02$\pm$0.01 & 4.44$\pm$0.30 \\
         & 500 & 3.95$\pm$0.04 & 1.99$\pm$0.01 & 4.30$\pm$0.27 \\
      \specialrule{.05em}{0.2em}{0.2em}
      27 & AD  & 4.02$\pm$0.04 & 2.10$\pm$0.02 & 4.67$\pm$0.22 \\
         & 500 & 4.01$\pm$0.05 & 2.09$\pm$0.02 & 4.59$\pm$0.19 \\
      \specialrule{.05em}{0.2em}{0.2em}
      40 & AD  & 4.09$\pm$0.06 & 2.23$\pm$0.02 & 5.02$\pm$0.18 \\
         & 500 & 4.07$\pm$0.07 & 2.19$\pm$0.02 & 4.80$\pm$0.16 \\
      \specialrule{.05em}{0.2em}{0.2em}
      48 & 500 & 4.06$\pm$0.06 & 2.25$\pm$0.02 & 4.97$\pm$0.19 \\
      \specialrule{.08em}{0em}{0em}
  \end{tabular*}
\end{table}

In Fig.~\ref{fig:bondangles}, we present bond-angle distributions computed from the structural models. These distributions provide a more detailed view of the local geometry than the coordination numbers alone, because they quantify how strongly the cation- and oxygen-centered polyhedra are distorted. The O--Ge--O bond angles, subtended at Ge centers, exhibit a peak at $\sim$106$\degree$, close to the tetrahedral geometry expected for GeO$_4$ units. For comparison, O--Ge--O bond-angle distributions in pure amorphous \germania{} have been reported to peak at 109.47$\degree$ \cite{neuefeind1996bond}, as indicated by the dashed line ($l_1$). The slight shift to lower angle and the broadening of the distribution indicate that the GeO$_4$ tetrahedra remain intact but become distorted by the surrounding Ti--O network.

With increasing Ti concentration, the O--Ge--O distributions broaden and decrease in peak height. This broadening reflects a wider range of Ge-centered tetrahedral distortions, which is expected when GeO$_4$ units are embedded in a chemically mixed network containing Ti-centered polyhedra of variable coordination. Notably, the distributions for the 40\% Ti-doped films are broader than those for the 48\% Ti-doped films, suggesting a non-monotonic structural evolution with Ti content. This behavior is consistent with the formation of more Ti-rich regions in the 48\% Ti-doped films, which leaves fewer strongly distorted Ge-centered polyhedra compared to films in which Ti is more intermixed with the Ge--O network.

The oxygen-centered bond-angle distributions, shown in the middle panel of Fig.~\ref{fig:bondangles}, are especially sensitive to how neighboring cation-centered polyhedra are connected. In a tetrahedral \germania\ network, the Ge--O--Ge angle is associated with corner-sharing GeO$_4$ tetrahedra and has been reported to peak near 133$\degree$ in amorphous \germania{} \cite{neuefeind1996bond}. In the present \tdg\ models, the oxygen-centered distributions include Ge--O--Ge, Ge--O--Ti, and Ti--O--Ti contributions. Their broad shape therefore reflects the coexistence of multiple types of bridging oxygen environments and the increasing role of Ti-centered polyhedra. As Ti concentration increases, the oxygen-centered distribution becomes less purely \germania-like and more influenced by Ti--O--Ti linkages, which in amorphous \titania\ are associated with characteristic angles near $\sim$102$\degree$ and $\sim$125$\degree$ \cite{triana2016electronic}.

The O--Ti--O bond-angle distributions, shown in the bottom panel, further illustrate the difference between Ge- and Ti-centered environments. Unlike Ge, which remains predominantly tetrahedral, Ti exhibits broader and more composition-dependent local geometry. The O--Ti--O distributions are broad and centered at substantially lower angles than the ideal tetrahedral angle, consistent with the higher and more variable Ti coordination indicated by Fig.~\ref{fig:coordination} and Table~\ref{table:2}. The reference O--Ti--O peak position near $\sim$94$\degree$ for amorphous \titania{} \cite{triana2016electronic}, marked by $l_5$, provides a useful comparison: with increasing Ti concentration, the Ti-centered environments in \tdg\ become increasingly \titania-like, although the films remain amorphous and structurally mixed. Together, the coordination and bond-angle results show that Ti incorporation preserves much of the Ge-centered tetrahedral backbone while progressively introducing higher-coordination, more distorted, and more compact Ti--O environments.

\subsection{Modeling: Intermediate Range Order}
\subsubsection{Polyhedral connections}
\begin{figure}[t]
	\centering
	\includegraphics[width=\linewidth]{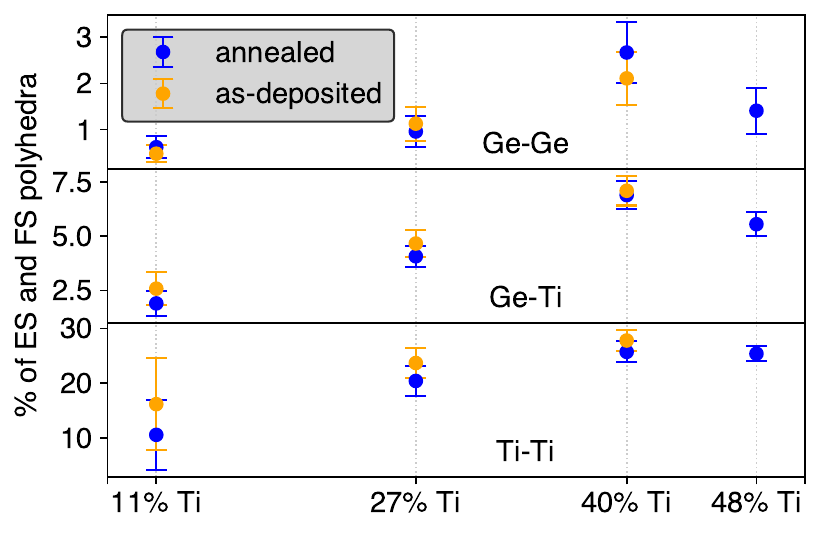}
	\caption{Percentage of edge-shared (ES) and face-shared (FS) polyhedra computed from the models. ``Ge--Ge'' refers to connections between two Ge-centered polyhedra. ``Ge--Ti'' refers to connections between a Ge-centered polyhedron and a Ti-centered polyhedron. ``Ti--Ti'' refers to connections between two Ti-centered polyhedra.}
	\label{fig:edge-face-shared}
\end{figure}

Looking beyond the first coordination sphere, we analyze polyhedral connections. All Ge- or Ti-centered polyhedra are connected to adjacent polyhedra through one or more shared O atoms at their vertices. Pairs of polyhedra sharing one, two, or three (or more) vertices are commonly referred to as corner-sharing (CS), edge-sharing (ES), or face-sharing (FS) polyhedra, respectively. In Fig.~\ref{fig:edge-face-shared}, we plot the percentage of ES and FS polyhedra computed from the models. The percentages of ES and FS polyhedra increase with Ti content, consistent with $n_{\mathrm{Ti}}>4$. Notably, the ES and FS fractions peak at 40\% Ti content, and models with 48\% Ti content exhibit lower ES and FS fractions than the 40\% Ti models. Since earlier work identified a correlation between the density of ES and FS motifs and mechanical loss \cite{prasai2019high}, these trends motivate closer attention to intermediate Ti concentrations when optimizing for low mechanical loss. 

Zhang \emph{et al.} used classical force-field MD simulations followed by DFT-based energy minimization to construct structural models of 44\% \tdg~\cite{zhang2026network}. They report the densities of CS, ES, and FS polyhedra to be 89\%, 10.6\%, and 0.4\%, respectively. For comparison, in the 40\% \tdg~models presented in this work, the corresponding densities of CS, ES, and FS polyhedra are 90.83$\pm$0.50\%, 8.67$\pm$0.49\%, and 0.5$\pm$0.13\%, respectively. These values are in close agreement within the stated uncertainties, taking into account the difference in Ti concentration between the two studies.

\subsubsection{Bond-orientational order around Ti}
We analyze the Steinhardt bond-orientational order parameters \cite{steinhardt1983bond}, focusing on $q_6$ for Ti atoms in the models, to assess the extent to which local Ti environments resemble those of common crystalline \titania~polymorphs. Here, $q_6$ is a rotationally invariant local order metric calculated from the angular distribution of bonds connecting each atom to its neighbors. In the normalization used here, $q_6$ takes values between 0 and 1. A higher value of $q_6$ for a given Ti atom indicates that its bond pattern projects more strongly onto sixfold angular symmetry. The precise numerical value of $q_6$ depends on how the neighbor shell is defined, such as the cutoff distance and/or the number of neighbors included; therefore, absolute values should be interpreted only within the same analysis protocol. For this analysis, all neighbors within a cutoff distance of 2.6~\AA\ are included.

\begin{figure}[t]
	\centering
	\includegraphics[width=\linewidth]{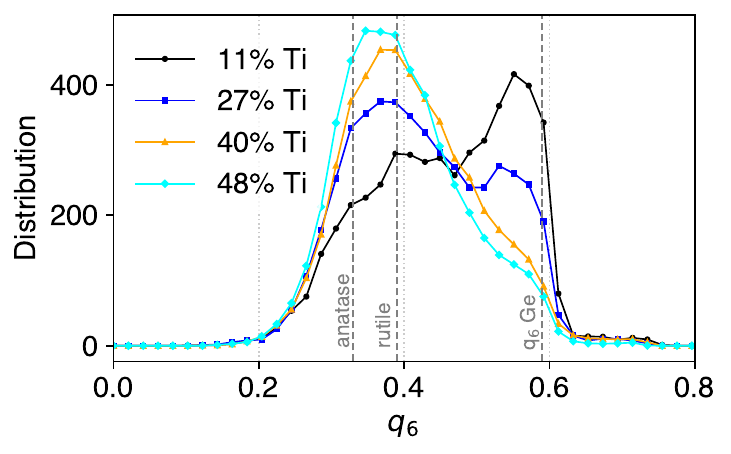}
	\caption{Normalized distribution of the Steinhardt bond-orientational order parameter $q_6$ (see text for details) for Ti atoms in the annealed models. For reference, the $q_6$ values of Ti atoms, computed using the same definition, in anatase and rutile structures are indicated by dashed lines. The peak position of the $q_6$ distribution for Ge atoms obtained from AIMD models is also indicated for comparison.}
	\label{fig:q6}
\end{figure}

The normalized distribution of $q_6$ is shown in Fig.~\ref{fig:q6}. Within the definition outlined above, Ti atoms in ideal crystalline \titania~polymorphs anatase and rutile exhibit $q_6$ values of 0.33 and 0.39, respectively. Ge atoms in AIMD models exhibit a $q_6$ distribution that peaks at 0.59. These reference values are indicated by dashed lines in Fig.~\ref{fig:q6}. At 11\% Ti concentration, Ti atoms exhibit a $q_6$ distribution that peaks at $q_6=0.56$, close to the peak position for Ge atoms. This observation, combined with the analysis of coordination numbers and bond-angle distributions discussed earlier, indicates that at low Ti concentrations, Ti atoms adopt a host-like local geometry. As the Ti concentration increases, the fraction of Ti atoms with anatase-like $q_6$ environments grows. Although the $q_6$ peak position is also close to that expected for rutile-like Ti, experimental studies show that these films crystallize into anatase upon annealing above 700$^\circ$C, supporting the assignment of these environments as anatase-like. The films measured in this work are fully amorphous, as demonstrated by the GIPDF data. Previous grazing-incidence x-ray diffraction studies have shown that \tdg~films (up to a Ti cation ratio of $\sim$50\%) remain amorphous up to annealing temperatures of 600$^\circ$C and show signatures of crystallization only at higher annealing temperatures \cite{vajente2021low}.

While Fig.~\ref{fig:q6} does not show a clear separation into distinct peaks, the systematic shift in the $q_6$ distribution indicates a progressive change in local Ti geometry with increasing Ti concentration. In particular, Ti coordination is predominantly tetrahedral at 11\% Ti, whereas it is almost entirely anatase-like at 40\% Ti. Although the density of anatase-like Ti increases overall with Ti concentration, this increase is especially sharp between the 27\% Ti and 40\% Ti films. This trend is consistent with the interpretation that, as Ti concentration increases, Ti atoms increasingly occupy Ti-rich local environments rather than isolated host-like sites. This suggests that the abrupt increase in anatase-like Ti environments between 27\% and 40\% Ti arises because the network provides a rapidly diminishing opportunity for Ti atoms to reside in small, isolated Ti clusters rather than larger anatase-like clusters as the Ti concentration increases.

\subsubsection{First Sharp Diffraction Peaks} \label{sec:fsdp}
\begin{figure}[h]
	\centering
	\includegraphics[width=\linewidth]{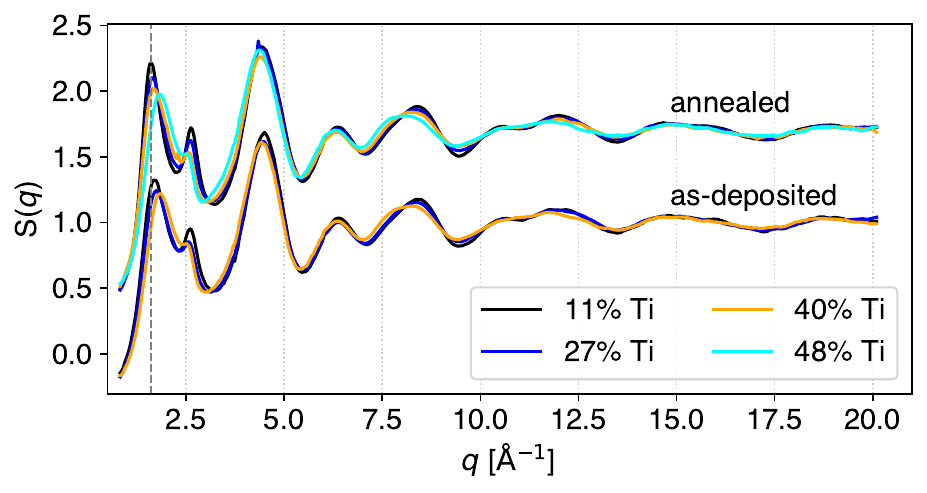}
	\caption{Measured structure factors $S(q)$ from thin films of \tdg. The $S(q)$ curves of annealed films are shifted along the vertical axis by 0.7 units for clarity. A dashed grey line marks 1.6~\AA$^{-1}$, near the first sharp diffraction peak (FSDP). The same data-processing protocol was used to obtain all curves. The observed differences are robust to small changes in the data-processing choices.}
	\label{fig:measured-Sq}
\end{figure}

Additional information on IRO can be obtained by analyzing the structure factor $S(q)$ (Fig.~\ref{fig:measured-Sq}). The first sharp diffraction peak (FSDP) is commonly used as a measure of IRO in amorphous networks. From Fig.~\ref{fig:measured-Sq}, the FSDP broadens and shifts to higher $q$ with increasing Ti content, indicating a reduction in the correlation length associated with the FSDP. Elliott, in Ref. \cite{elliott1991origin}, attributed the FSDP in \silica~and other AX$_2$-type glasses to changes in atom-sized interstitial voids surrounding cation-centered polyhedral units. This picture is consistent with our structural analysis: Ti addition increases the average coordination and the fraction of compact edge-sharing polyhedra, which is consistent with a reduction of open space surrounding the polyhedral units. Prior high-pressure studies of amorphous \silica\ and \germania\ show that densification shifts the FSDP to higher $q$ and modifies its intensity/width, consistent with a reduction of intermediate-range correlation length and collapse of open ring/void topology~\cite{sampath2003intermediate, guthrie2004formation, benmore2010structural}. Other works on silica and silicate glasses has shown that changes in FSDP intensity and width are linked to changes in ring-size distributions, cage geometry, and the periodicity of ring-centered structural correlations \cite{guerette2015structure,shi2019ring,kobayashi2023machine}.

Upon annealing, however, we observe that the FSDP sharpens, increases in intensity, and shifts to lower $q$, although these effects are subtle. This suggests that annealing increases the interstitial volume around Ti-centered polyhedra, consistent with reduced fractions of ES polyhedra.

\subsection{Structure Correlation with Mechanical Loss}
The samples for which GIPDF data were obtained were also measured for mechanical loss (Table~\ref{table:1}). In this section, we analyze whether correlations exist between structural features and the measured mechanical loss. We tested a number of structural descriptors for possible correlations with mechanical loss, including short-range features such as coordination statistics and bond-angle distributions, as well as intermediate-range features such as bond order, polyhedral connections, and the positions and shapes of the FSDPs. Since there are only seven samples with both GIPDF and mechanical loss data, any observed correlation is necessarily limited in statistical strength and should therefore be interpreted with caution.

\begin{figure}[h]
	\centering
	\includegraphics[width=\linewidth]{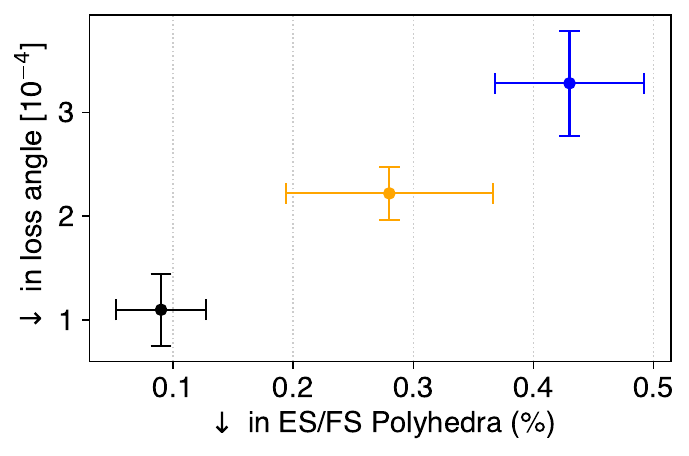}
	\caption{Effect of annealing on mechanical loss and polyhedral connectivity. The annealing-induced decrease in mechanical loss is plotted against the corresponding decrease in the fraction of edge-sharing (ES) and face-sharing (FS) polyhedra. Each point represents one composition for which both as-deposited and annealed GIPDF/model data are available. Colors denote Ti concentration: black for 11\% Ti, blue for 27\% Ti, and orange for 40\% Ti.}
	\label{fig:annealing-correlation}
\end{figure}

For the three compositions with paired as-deposited and annealed GIPDF data, the clearest trend is observed between the decrease in mechanical loss upon annealing and the decrease in the fraction of ES and FS polyhedra. It is important to note that the observed correlation is with the \emph{change}, rather than the absolute value, in the density of ES and FS polyhedra. Even though the annealing-induced decrease in the density of ES and FS polyhedra is rather small, these changes exhibit a strong correlation with the decrease in mechanical loss (Pearson correlation coefficient is 1.0 with a $p$-value of 0.03), see Fig.~\ref{fig:annealing-correlation}. This suggests that only a minority of the ES and FS polyhedra contribute strongly to the mechanical loss, and that these are preferentially reduced by annealing. In terms of the TLS framework, this picture would be consistent with the highest-energy barrier TLSs being removed first during annealing and also being the dominant contributors to mechanical loss. Empirically it has been seen that annealing 44\% \tdg\ at 600$^\circ$C for more than 100 hours results in much larger changes in mechanical loss than 10 hours at 500$^\circ$C \cite{vajente2021low}. Future work comparing GIPDF data for such layers would provide useful information on this point. A correlation between room-temperature mechanical loss and the density of ES and FS polyhedra was first reported in Ref.~\cite{prasai2019high}.

Although we tested several additional structural descriptors, none showed a correlation with mechanical loss comparable to the annealing-induced change in edge- and face-sharing polyhedra. Among the remaining descriptors, the peak height of the FSDP shows the strongest trend with room-temperature mechanical loss, with Pearson correlation coefficient $-0.6$ and p-value 0.16. This correlation is not statistically strong, but its sign is physically meaningful: because the FSDP reflects intermediate-range correlations in the amorphous network, a larger FSDP height is consistent with stronger IRO and a more relaxed polyhedral packing. Thus, the FSDP trend provides secondary evidence that improved intermediate-range organization may be favorable for lower mechanical loss. However, the unusually low loss of the 27\% annealed sample indicates that mechanical loss is not controlled by a single structural descriptor; composition-dependent changes in local Ti environments and polyhedral connectivity likely also play important roles.

\begin{figure}[t]
	\centering
	\includegraphics[width=\linewidth]{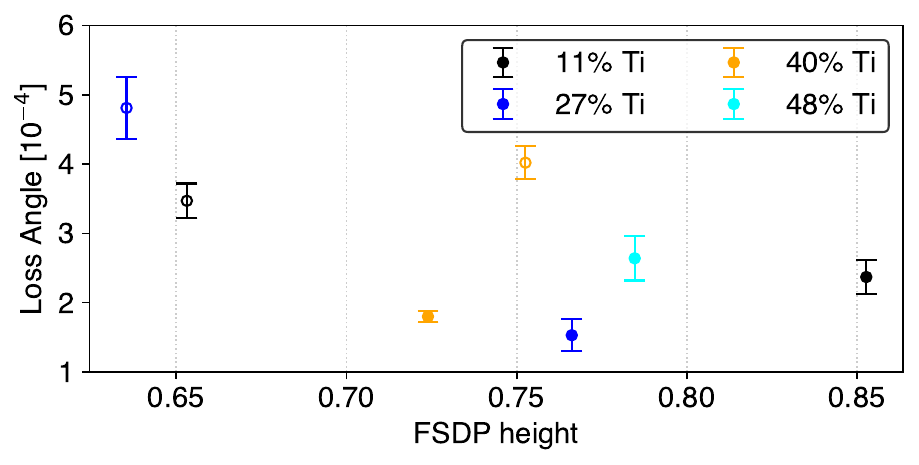}
	\caption{Correlation between the measured loss angle and the height of the first sharp diffraction peak (FSDP). Filled circles represent annealed samples, open circles represent as-deposited samples. Pearson correlation coefficient between FSDP height and loss angle is -0.6 with a p-value of 0.16.}
	\label{fig:fsdp-correlation}
\end{figure}

\section{CONCLUSIONS}

We have combined grazing-incidence X-ray pair distribution function measurements with experimentally constrained atomic-structure modeling to determine how Ti concentration and annealing modify the atomic structure of amorphous TiO$_2$-doped GeO$_2$ optical coatings. The central result is that Ti doping does not simply perturb a GeO$_2$ network in a uniform way. Rather, increasing Ti concentration progressively changes the short- and intermediate-range network structure: Ti coordination increases, Ti-rich local environments become more prevalent, and compact edge- and face-sharing polyhedral connections grow in density, while Ge remains predominantly fourfold coordinated. Annealing drives the structure in the opposite direction -- it reduces the fraction of edge- and face-sharing polyhedra and sharpens the first sharp diffraction peak, consistent with a more relaxed intermediate-range structure. 

Among the structural descriptors examined here, the reduction in mechanical loss upon annealing is most strongly correlated with the decrease in edge- and face-sharing polyhedra. A weaker, but physically consistent, trend between lower mechanical loss and larger FSDP height further suggests that enhanced intermediate-range order is favorable for reduced dissipation. Taken together, these observations suggest that room-temperature mechanical loss in TiO$_2$-doped GeO$_2$ is closely linked to the population and relaxation of specific intermediate-range motifs---particularly compact shared-polyhedra connections---whose abundance is set by both Ti concentration and thermal processing.

These results provide an atomic-scale picture of why TiO$_2$-doped GeO$_2$ is a promising high-index coating material for gravitational-wave detectors, while also clarifying the structural tradeoff involved in its optimization. Higher Ti content can increase refractive index and reduce coating thickness, but it also promotes Ti-rich, compact polyhedral environments that may increase mechanical loss. The most useful design principle suggested by this work is therefore not simply to maximize or minimize Ti concentration, but to identify compositions and processing conditions that preserve the optical advantages of Ti doping while suppressing mechanically lossy shared-polyhedra motifs and enhancing relaxed intermediate-range order.

Finally, the combined experimental and modeling framework developed here provides a route for connecting coating synthesis, amorphous structure, and mechanical dissipation in a quantitatively testable way. Extending this approach to additional compositions, annealing schedules, and deposition conditions should help identify coating structures with lower thermal noise for future gravitational-wave detectors and other precision optical systems.

\section{DATA AVAILABILITY}
Data are available from the corresponding author upon reasonable request.

\section{ACKNOWLEDGMENTS}
This work was supported by the Gordon and Betty Moore Foundation, grant DOI 10.37807/GBMF6793.02. MMF acknowledges support from National Science Foundation (NSF), grant numbers 2309289  and NSF 2309086. KHL acknowledges support by Research Foundation of Korea (NRF) grant funded by the Korean government (MSIT) (No. RS-2024-00455482). RBS analysis performed at U. Montréal was possible thanks to funding from the CFI, the NSERC and the FRQNT through the RQMP. This paper has LIGO document No. P2600195.

\appendix
\section{Classical Force Field} \label{app:potential}
The molecular dynamics and structure relaxations in this work used a classical force field as described below. 

The total force field was written as

\begin{equation}
U_{\rm FF}
=
U_{\rm TiGeO}^{\rm BKS+Morse+3b}
+
U_{\rm sr}
+
U_{\rm Ar-Ar}
+
U_{\rm Ar-host}.
\end{equation}
The main Ti--Ge--O network interaction was described using a BKS--Morse force field with an added harmonic three-body term, following Ref.~\cite{jiang2025machine, trinastic2013unified}. To keep the notation consistent with the LAMMPS implementation, we write
\begin{equation}
U_{\rm TiGeO}^{\rm BKS+Morse+3b}
=
U_{\rm BKS}
+
U_{\rm Morse}
+
U_{\rm 3b},
\label{eq:forcefield}
\end{equation}
where
\begin{align}
U_{\rm BKS}
=&
\sum_{i<j}
\left[
\frac{k_e q_i q_j}{r_{ij}}
+
A_{ij}\exp\left(-\frac{r_{ij}}{\rho_{ij}}\right)
-
\frac{C_{ij}}{r_{ij}^{6}}
\right],
\\
U_{\rm Morse}
=&
\sum_{i<j}^{\rm Ti-O}
D_{ij}
\left[
e^{-2\alpha_{ij}(r_{ij}-r^0_{ij})}
-
2e^{-\alpha_{ij}(r_{ij}-r^0_{ij})}
\right],
\\
U_{\rm 3b}
=&
\sum_{\angle j-i-k}
K_{jik}
\left(\theta_{jik}-\theta^0_{jik}\right)^2 .
\end{align}
Here \(q_{\rm Ge}=+2.4\), \(q_{\rm Ti}=+2.4\), \(q_{\rm O}=-1.2\), and \(q_{\rm Ar}=0\). Long-range Coulomb interactions were evaluated using Ewald summation with a tolerance of \(10^{-6}\). The BKS and Morse parameters are given in Table~\ref{table:forcefield}. Cation--cation pairs were included through the Coulomb part of the interaction, with zero short-range Buckingham parameters.

\begin{table}[h]
\caption{BKS and Morse two-body force-field parameters used for the oxide network. Blank entries indicate that the corresponding term was not applied for that pair.}
\label{table:forcefield}
\centering
\footnotesize
\setlength{\tabcolsep}{3.5pt}
\renewcommand{\arraystretch}{1.15}
\begin{tabular}{lcccccc}
\hline
pair
& \(A\)
& \(\rho\)
& \(C\)
& \(D\)
& \(\alpha\)
& \(r^0\) \\
& (eV)
& (\AA)
& (eV \AA$^6$)
& (eV)
& (\AA$^{-1}$)
& (\AA) \\
\hline
Ge--O
& 42206.1
& 0.1605
& 0.00838283
& --
& --
& -- \\
O--O
& 1388.773
& 0.3623188
& 175.0
& --
& --
& -- \\
Ti--O
& 5505.120
& 0.2253
& 20.0
& 0.5478
& 1.9
& 1.96 \\
\hline
\end{tabular}
\end{table}

The only nonzero three-body parameter was for O-centered Ge--O--Ge triplets, with \(K_{\rm GeOGe}=3.00\) and \(\theta^0_{\rm GeOGe}=120^\circ\), using an O--Ge neighbor cutoff of 2.20~\AA. All other three-body entries were set to zero.

A short-range term was added to selected oxide-network pairs to prevent unphysical close approaches during high-temperature melting and refinement:
\begin{equation}
U_{\rm sr}(r)
=
4\epsilon_{ij}
\left[
\left(\frac{\sigma_{ij}}{r}\right)^{24}
-
\left(\frac{\sigma_{ij}}{r}\right)^6
\right].
\end{equation}
This term was applied with a cutoff of 10.0~\AA. The parameters were \(\epsilon=0.000357\)~eV and \(\sigma=1.900\)~\AA{} for O--O pairs, and \(\epsilon=0.010\)~eV and \(\sigma=1.000\)~\AA{} for Ti--O pairs.

The Ar--Ar interaction was described using
\begin{equation}
U_{\rm Ar-Ar}(r)
=
4\epsilon_{\rm Ar-Ar}
\left[
\left(\frac{\sigma_{\rm Ar-Ar}}{r}\right)^{12}
-
\left(\frac{\sigma_{\rm Ar-Ar}}{r}\right)^6
\right],
\end{equation}
with \(\epsilon_{\rm Ar-Ar}=0.0104\)~eV, \(\sigma_{\rm Ar-Ar}=3.40\)~\AA, and a cutoff of 8.5~\AA~\cite{bernardes1958theory}. Interactions between Ar and the oxide-network atoms were modeled using the ZBL potential~\cite{ziegler1985stopping},
\begin{equation}
U_{\rm Ar-host}(r)
=
S(r)
\frac{k_e Z_i Z_j}{r}
\Phi\left(\frac{r}{a}\right),
\end{equation}
where \(S(r)\) is the switching function for the inner and outer cutoffs of 5.0 and 8.5~\AA,
\begin{equation}
a=
\frac{0.46850}{Z_i^{0.23}+Z_j^{0.23}},
\end{equation}
and
\begin{align}
\Phi(x)
=&\,
0.1818e^{-3.2x}
+0.5099e^{-0.9423x} \nonumber \\
&+
0.2802e^{-0.4029x}
+0.02817e^{-0.2016x}.
\end{align}
The atomic numbers used for the ZBL terms were \(Z_{\rm Ge}=32\), \(Z_{\rm O}=8\), \(Z_{\rm Ti}=22\), and \(Z_{\rm Ar}=18\).

\section{Comparison with AIMD Models} \label{app:aimd}

To provide an independent first-principles reference for the atomic structure, we generated 30\% \tdg\ models using an \textit{ab initio} molecular dynamics (AIMD) melt--quench workflow \cite{drabold2009topics}. This composition lies within the measured 11--48\% Ti range and serves as an intermediate reference for comparison with the experimentally constrained models.

The AIMD simulations were carried out using the Vienna \textit{Ab initio} Simulation Package (VASP) \cite{kresse1993ab,kresse1996efficient}. Electron--ion interactions were described using the projector augmented-wave method \cite{blochl1994projector}, and exchange and correlation were treated with the PBE functional \cite{perdew1996generalized}. The valence states were expanded in a plane-wave basis with a kinetic-energy cutoff of 300~eV. Because of the size of the amorphous supercells, Brillouin-zone sampling was restricted to the $\Gamma$ point. A time step of 1~fs was used throughout the simulations.

\begin{figure}[t]
	\centering
	\includegraphics[width=\linewidth]{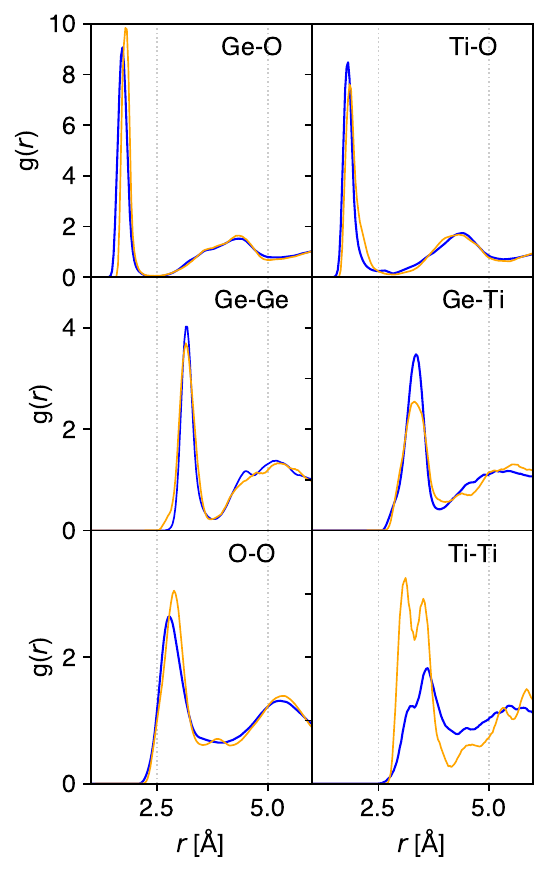}
	\caption{Comparison of partial PDFs computed from the experimentally constrained models in this work and from AIMD models. Orange curves are computed from 30\%~\tdg\ AIMD models generated using the protocol described in this section. Blue curves are computed from 27\% Ti models (annealed) presented in this paper. To reduce noise, all curves are smoothed using a running-window average with a window size of six data points.}
	\label{fig:partials}
\end{figure}

The starting configuration was a 360-atom model of amorphous \germania\ from Ref.~\cite{prasai2023glass}. 30\% of the Ge atoms were replaced at random by Ti atoms, and the simulation cell was rescaled to an initial density of 3.65~g/cm$^3$, close to the measured film densities listed in Table~\ref{table:1}. The preparation was carried out in two stages. In the first stage, the model was equilibrated at 800~K for 1~ps, cooled to 300~K over 2~ps, and then equilibrated at 300~K for 5~ps. Five snapshots from the final 300~K trajectory were subjected to simultaneous ionic and cell-volume relaxation, yielding an average relaxed density of 3.77~g/cm$^3$. In the second stage, the final 300~K configuration was rescaled to this density, equilibrated at 2000~K for 10~ps, cooled to 300~K over 10~ps, and equilibrated at 300~K for an additional 10~ps. One hundred snapshots from this final 300~K trajectory were used to compute the AIMD reference partial PDFs shown in Fig.~\ref{fig:partials}.
For comparison with the AIMD reference, we plot the partial PDFs from the 27\% Ti annealed models presented in this work, since this composition is closest to the 30\% AIMD composition. 

Overall, the partial PDFs from the experimentally constrained models compare reasonably well with the AIMD reference. The main exception is the Ti--Ti partial PDF, which is more disordered in the experimentally constrained models than in the AIMD models. The origin of this difference is not fully resolved. One possible contribution is the smaller cell size of the AIMD models compared with the experimentally constrained models used in this work. In finite amorphous simulation cells, periodic boundary conditions and limited configurational sampling can introduce finite-size artifacts, including spuriously enhanced intermediate- or long-range order and incomplete sampling of local structural motifs \cite{nakano1994first,mora2019making}. A second possible contribution is the explicit inclusion of Ar atoms in the experimentally constrained models. Because Ar was detected in the measured films, its inclusion makes those models closer to the experimental film composition and may introduce structural disorder that is absent from the AIMD reference. Finally, part of the difference may reflect limitations of the BKS--Morse force field in describing Ti--Ti correlations during the refinement procedure. The Ti--Ti partial PDFs are also noisier at lower Ti concentrations because fewer Ti--Ti pairs contribute to the statistics.

\bibliography{kbib}
\bibliographystyle{apsrev4-1}

\end{document}